\documentclass{webofc}
\usepackage[varg]{txfonts}   
\usepackage{lineno}
\begin{document}
%
\title{Anisotropic Flow of Identified Particles in Au + Au Collisions at $\sqrt{s_{NN}}$ = 3 - 3.9 GeV at RHIC}
%
%

\author{\firstname{Zuowen} \lastname{Liu}\inst{1}\fnsep\thanks{\email{liuzw@mails.ccnu.edu.cn}}
        ~for the STAR Collaboration
}

\institute{Central China Normal University, Wuhan, Hubei 430079 }

\abstract{
In these proceedings, we present transverse momentum dependence of the mid-rapidity slope of directed flow ($dv_1/dy|_{y=0}$) for $\pi^+$ and $K_S^0$
in Au + Au collisions at $\sqrt{s_{NN}}$ = 3.0, 3.2, 3.5, and 3.9 GeV. Both $\pi^+$ and $K_S^0$ show negative $v_1$ slope at low $p_T$ ($p_T < 0.6$ GeV/$c$).
Collision energy dependence of $v_1$ slope and $p_T$-integrated $v_2$ for $\pi^{\pm}$, $K_S^0$, and $\Lambda$ are also presented.
A comparison to JAM model calculations indicates that spectator shadowing can lead to anti-flow at low $p_T$.
In addition, a breaking of the Number of Constitute Quark (NCQ) scaling of elliptic flow ($v_2$) is observed at $\sqrt{s_{NN}}$ = 3.2 GeV,
which implies the dominance of hadronic degrees of freedom occurs in collisions at $\sqrt{s_{NN}}$ = 3.2 GeV and below. 
}
 
\maketitle
\section{Introduction}
\label{intro}
The goals of Beam Energy Scan (BES) program at Relativistic Heavy Ion Collider (RHIC) are searching for the possible QCD critical point and locating the first order phase boundary~\cite{Luo:2020pef}.
The energy dependence of net-proton $v_1$ slope~\cite{PhysRevLett.112.162301} shows possible minimum at $\sqrt{s_{NN}} \approx$ 10 - 20 GeV, implies that the softest point of Equation of State (EoS) may exist within this range of collision energy.
The existence of partonic collectivity is observed through NCQ scaling of $v_2$ at higher BES energies ($\sqrt{s_{NN}} > $ 7.7 GeV)~\cite{PhysRevC.88.014902},
while the break NCQ scaling of $v_2$ at $\sqrt{s_{NN}}$ = 3.0 GeV~\cite{2022137003} indicates the partonic collectivity is disappeared at this energy.
In this contribution, we present the most recent measurements of directed flow ($v_1$) and elliptic flow ($v_2$) of identified particles ($\pi^\pm$, $K^{\pm}$, $K_S^0$, $p$, and $\Lambda$)
at $\sqrt{s_{NN}}$ = 3.0 - 3.9 GeV in Fixed Target Au + Au collisions.

\section{Experiment Setup}
\label{exper}
\begin{figure}[h]
\centering
\sidecaption
\includegraphics[width=0.35\linewidth,clip]{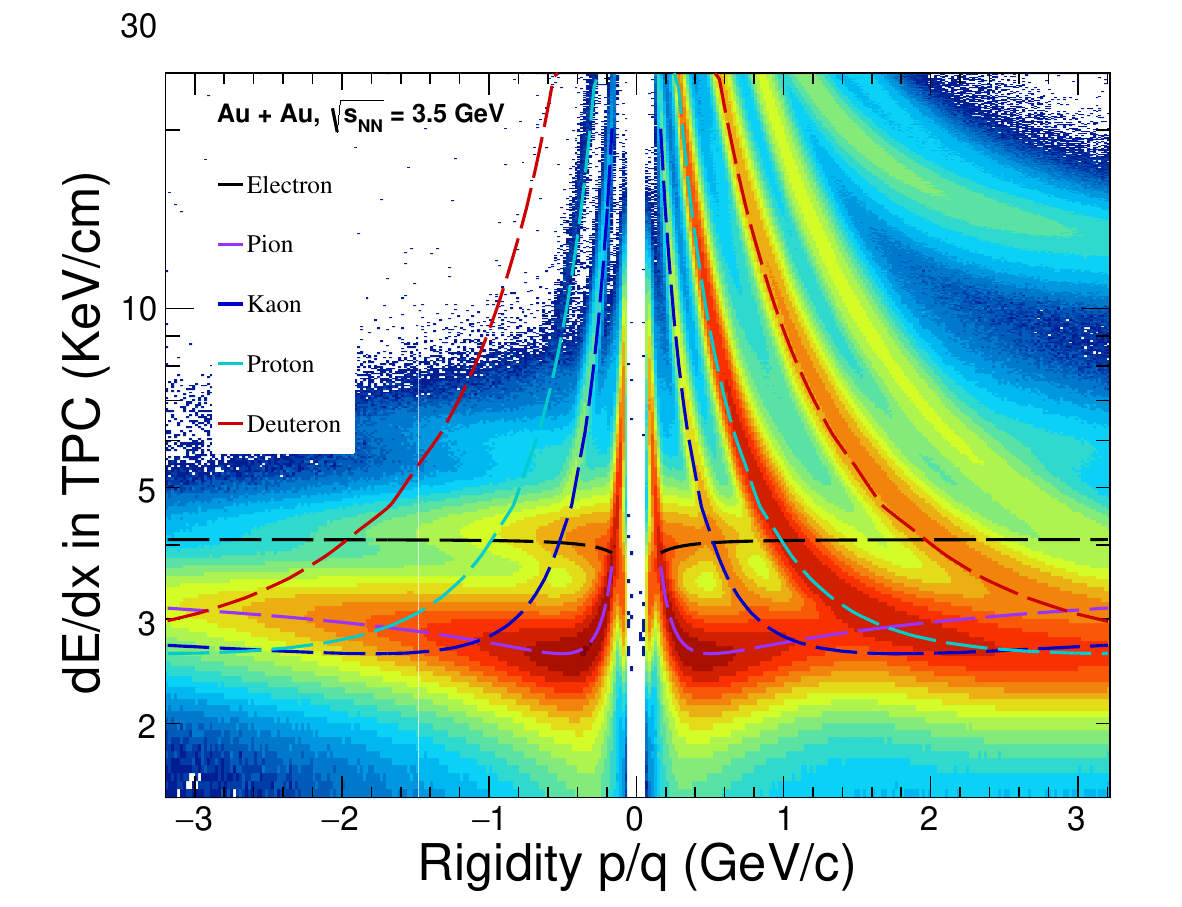}
\includegraphics[width=0.38\linewidth,clip]{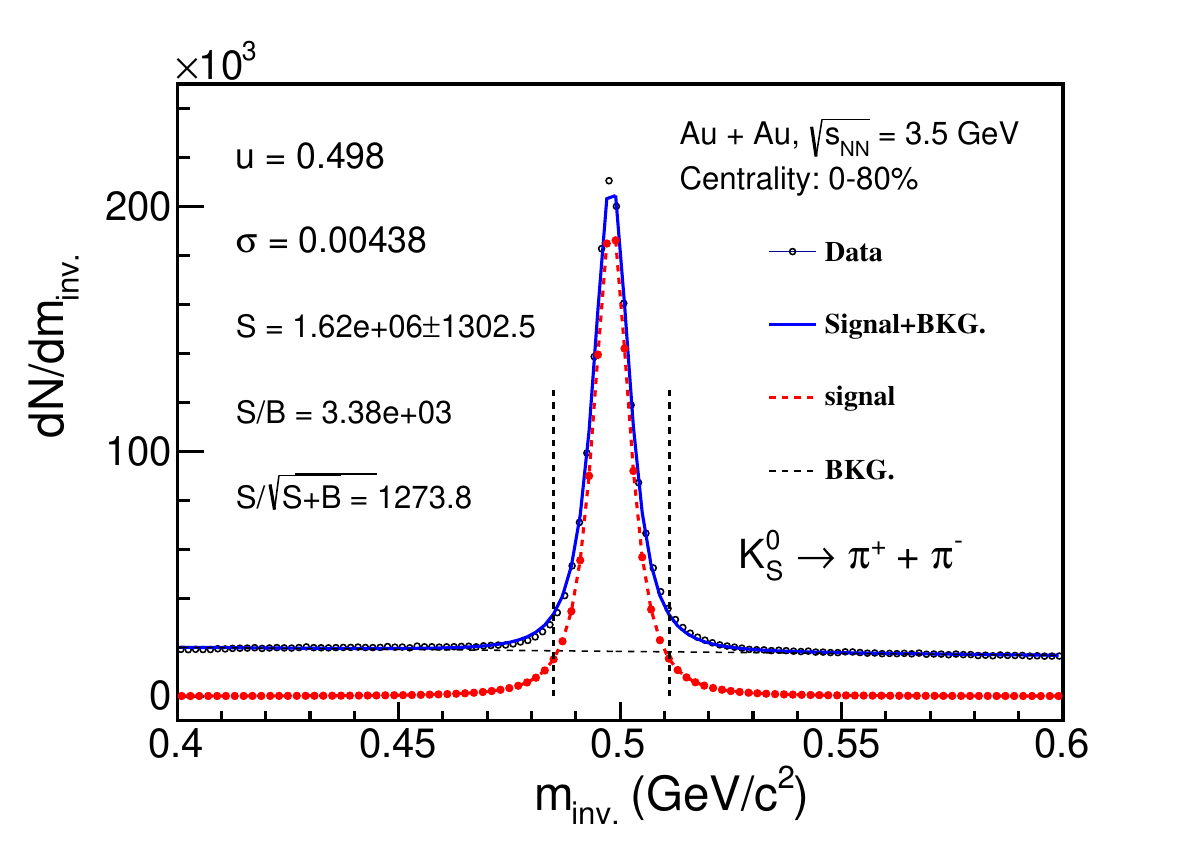}
\caption{Left: Rigidity dependence of particle ionization energy loss in TPC. Right: Invariant mass distribution of $K_S^0$ in Au + Au collisions at $\sqrt{s_{NN}}$ = 3.5 GeV.}
\label{fig:pid_plot}
\end{figure}

For identification of $\pi^{\pm}$, $K^{\pm}$, protons and anti-protons, a combination of Time Projection Chamber (TPC)~\cite{Anderson:2003ur} and Time of Flight (TOF)~\cite{Llope:2005yw} is used.
The left panel of Figure~\ref{fig:pid_plot} illustrates the rigidity ($p/q$: particle momentum divided by charge) dependence of ionization energy loss ($dE/dx$) in the TPC.
The dashed line represents the theoretical ionization energy loss curve for particle passing through the TPC. 
Particle identification by TOF is based on particle mass square ($m^2$) distribution, which can be obtained from particle velocity ($\beta$).
Moreover, the Kalman Filter (KF) particle package~\cite{Banerjee:2020iab}, where the covariance matrix of reconstructed tracks is taken into account,
is employed to reconstruct weak decay particles ($K_S^0$ and $\Lambda$). 
An example of reconstructing the invariant mass using the KF particle package is demonstrated with the $K_S^0$ meson in the right panel of Figure~\ref{fig:pid_plot}.

\section{Results}
\label{rsts}

\subsection{Anti-flow of Kaon}
\label{kaonAntiflow}

\begin{figure}[h]
\centering
\sidecaption
\includegraphics[width=0.65\linewidth,clip]{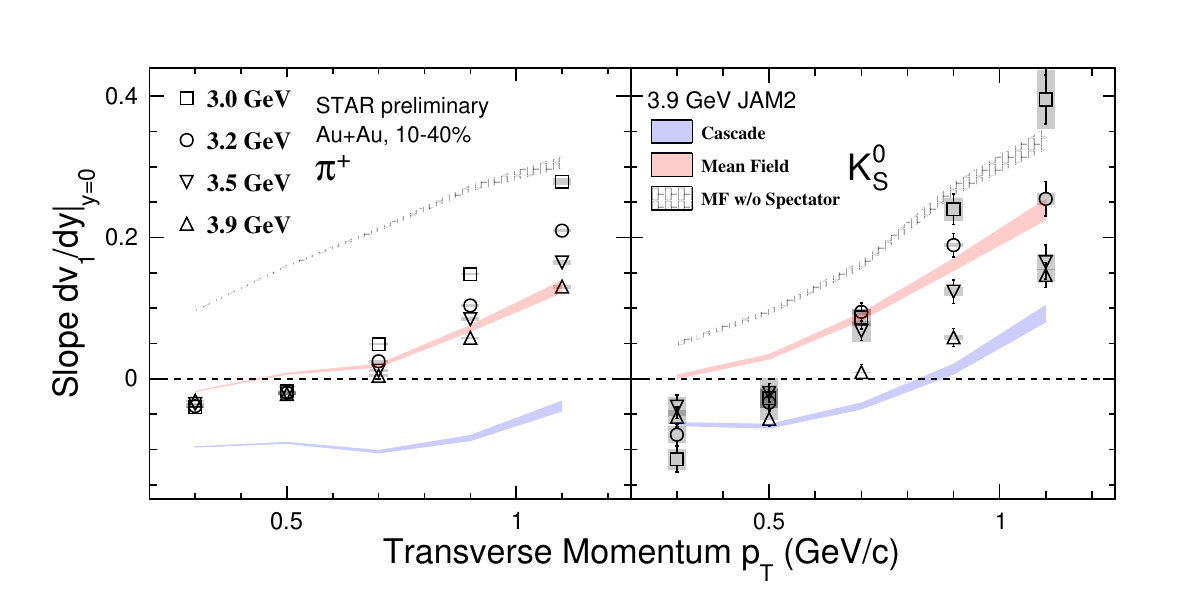}
\caption{$v_1$ slope of $\pi^+$(left) and $K_S^0$(right) as function of transverse momentum and a comparison with JAM calculation at $\sqrt{s_{NN}}$ = 3.9 GeV.}
\label{fig:anti-flow}
\end{figure}

The anti-flow of kaon was first observed by E895 Collaboration at 6 A GeV~\cite{PhysRevLett.85.940}. 
It was attributed to the repulsive potential associated with the strange quark in $K_S^0$.
We have observed anti-flow behavior in kaons and pions for $p_T$ < 0.6 GeV/$c$ in mid-central Au + Au collisions at $\sqrt{s_{NN}}$ = 3.0, 3.2, 3.5, and 3.9 GeV using the fixed target data from STAR. 
Figure~\ref{fig:anti-flow} shows transverse momentum ($p_T$) dependence of $v_1$ slope ($dv_1/dy|_{y=0}$) for $\pi^+$ and $K_S^0$ from STAR.
The hadronic transport model JAM~\cite{PhysRevC.105.014911} calculations are compared with experimental data at 3.9 GeV.
The JAM model in hadronic cascade mode (blue band) can successfully capture the anti-flow pattern at low $p_T$ for $\pi^+$ and $K_S^0$, 
even without the inclusion of a kaon potential~\cite{PhysRevLett.85.940}.
However, the JAM model with baryonic mean field (red band), tends to overestimate the $v_1$ slope for $\pi^+$ and $K_S^0$.
Additionally, the JAM mean field without spectator contribution (black band) exhibits a larger $v_1$ slope compared to the one with spectators. 
The data-model comparisons suggest that the shadowing effect~\cite{PhysRevC.59.348} from the spectator may also play a significant role in generating anti-flow at low $p_T$.

\subsection{NCQ Scaling of $v_2$}
\label{ncqScaling}
\begin{figure}[h]
\centering
\sidecaption
\includegraphics[width=0.65\linewidth,clip]{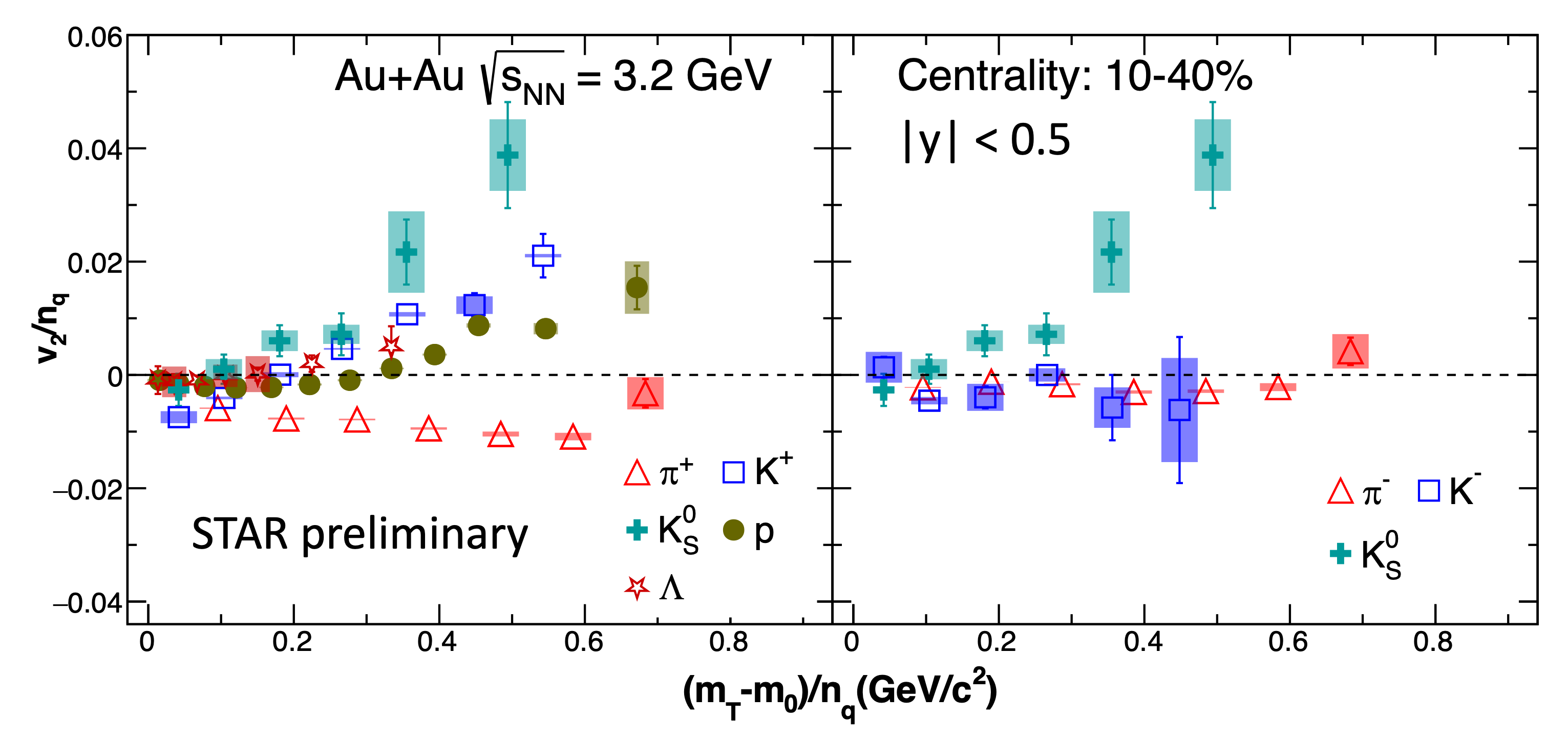}
\caption{Number of Constituent Quark scaling of $v_2$ as a function of the scaled transverse kinetic energy for particles (left) and the corresponding anti-particles (right) at $\sqrt{s_{NN}}$ = 3.2 GeV.} 
\label{fig:NCQ3p2}
\end{figure}

The left and right panels of Figure~\ref{fig:NCQ3p2} illustrate the number of constituent
quarks ($n_q$) scaled elliptic flow ($v_2/n_q$) as a function of transverse kinetic energy ($(m_T - m_0)/n_q$)
for particles ($\pi^+, K^+, K^0_S, p,$ and $\Lambda$) and the corresponding anti-particles ($\pi^-, K^-,$ and $K^0_S$), respectively for Au + Au collisions at $\sqrt{s_{NN}}$ = 3.2 GeV.
The NCQ scaling of $v_2$ is broken completely for particles and anti-particles at $\sqrt{s_{NN}}$ = 3.2 GeV.
The existence of partonic collectivity is observed through NCQ scaling of $v_2$ at higher BES energies ($\sqrt{s_{NN}} >$ 7.7 GeV)~\cite{PhysRevC.88.014902}.
The disappearing of NCQ scaling in $v_2$ at $\sqrt{s_{NN}}$ = 3.2 GeV implies that hadronic interactions play an important role at this energy and below~\cite{2022137003, Lan:2022rrc}.

\subsection{Energy Dependence of $v_1$ and $v_2$}
\label{v1v2energy}

\begin{figure}[h]
\centering
\sidecaption
\includegraphics[width=0.45\linewidth,clip]{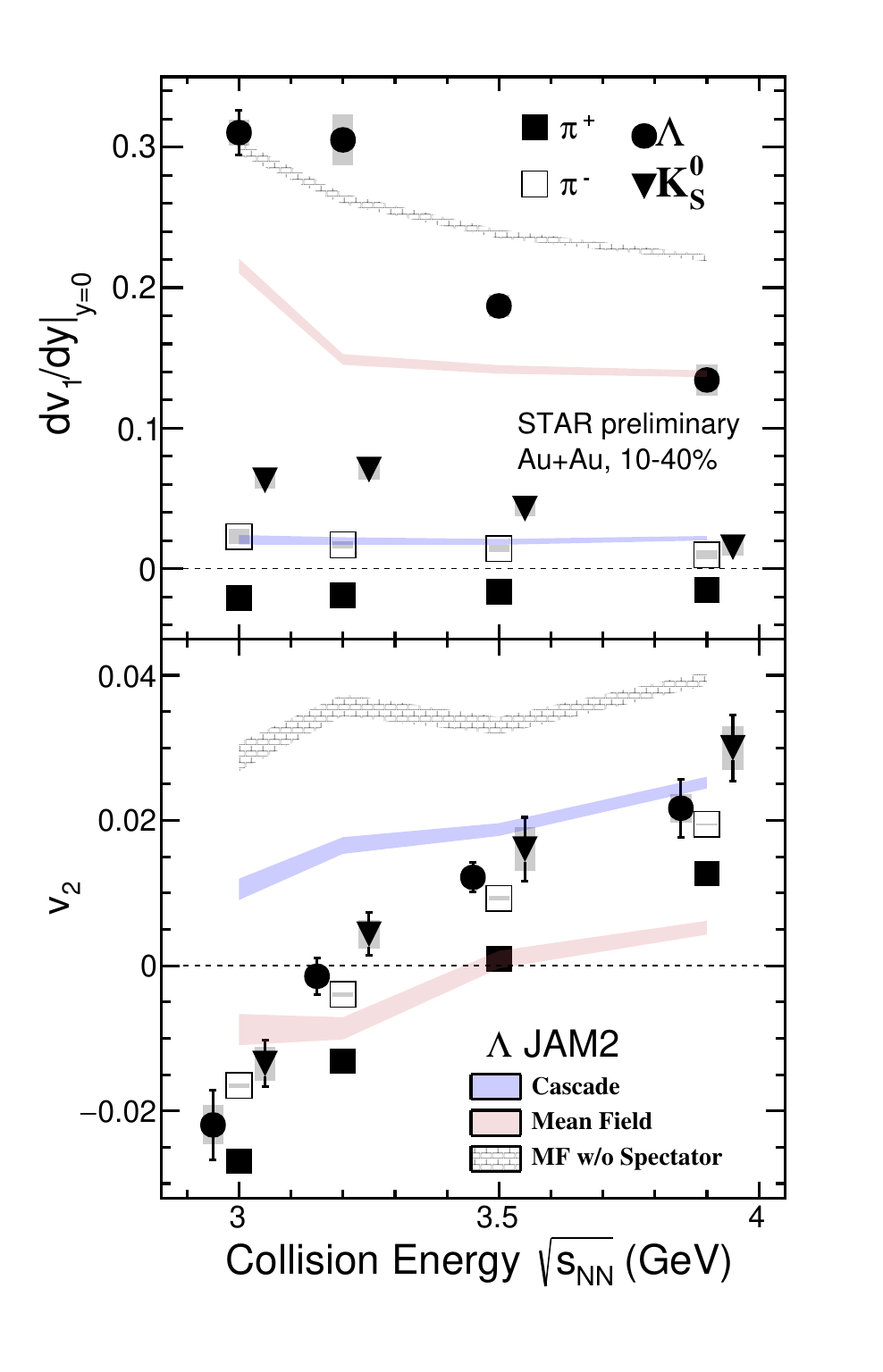}
\caption{$v_1$ slope (top) and $p_T$-integrated $v_2$ (bottom) as a function of collision energy and compared with JAM calculation for $\Lambda$.
Note that $p_T$ windows for $\pi^{\pm}$, $K_S^0$, and $\Lambda$ are $0.2 < p_T < 1.6$ GeV/$c$, $0.4 < p_T < 1.6$ GeV/$c$, and $0.4 < p_T < 2.0$ GeV/$c$, respectively.
And the rapidity window is $-0.5 < y < 0$ for $p_T$-integrated $v_2$.}
\label{fig:v1v2_E}
\end{figure}

The top panel of Figure~\ref{fig:v1v2_E} shows collision energy dependence of $v_1$ slope in 10-40\% mid-central Au + Au collisions at $\sqrt{s_{NN}}$ = 3.0 - 3.9 GeV. 
The $v_1$ slopes of $\pi^{+}$ (solid square) are negative at 3.0 - 3.9 GeV, while $v_1$ slopes of $\pi^{-}$ (open square) are positive. 
The difference between $\pi^{+}$ and $\pi^{-}$ may be explained by Coulomb effect~\cite{PhysRevC.98.055201}. 
Furthermore, $v_1$ slopes of $K_S^0$ (solid triangle) are greater than $\pi^{+}$, $v_1$ slopes of $\Lambda$ (solid circle) are largest among these four particle species. 
The $v_1$ slopes of all particles ($\pi^{\pm}, K_S^{0}$, and $\Lambda$) decrease in magnitude as collision energy increases.
The lower panel in Figure~\ref{fig:v1v2_E} depicts the collision energy dependence of transverse momentum ($p_T$) integrated $v_2$. 
It is observed that the sign of $v_2$ changes from negative to positive for all particles ($\pi^{\pm}, K_S^{0}$, and $\Lambda$)
within the collision energy range of $\sqrt{s_{NN}}$ = 3.0 - 3.9 GeV. This shift signifies the transition from out-of-plane to in-plane expansion~\cite{PhysRevLett.83.1295},
occurring specifically within the mentioned energy range.

The JAM calculations for $\Lambda$ are represented by colored bands, 
with blue, red, and black bands corresponding to cascade, baryonic mean field, and mean field without spectators modes, respectively.
The comparison of $\Lambda$ $v_1$ between the data and model calculations suggests 
the presence of a strong baryon mean field~\cite{2022137003} in the high baryon density region.
Furthermore, the comparison between the measured $p_T$-integrated $v_2$ and model calculations indicates
the significant influence of spectator shadowing in the energy range of $\sqrt{s_{NN}}$ = 3.0 - 3.9 GeV.

\section*{Summary}
\label{summ}
In summary, we present directed flow ($v_1$) and elliptic flow ($v_2$) measurements for identified particles ($\pi^\pm$, $K^{\pm}$, $K_S^0$, $p$, and $\Lambda$) in Au + Au collisions at $\sqrt{s_{NN}}$ = 3.0, 3.2, 3.5, and 3.9 GeV. 
The measurements for $\pi^+$ and $K_S^0$ show negative $v_1$ slope ($dv_1/dy|_{y=0}$) at low $p_T$ ($p_T < 0.6$ GeV/$c$).
The transport model JAM reproduces anti-flow at low $p_T$ without incorporating kaon potential, and indicates shadowing effect from spectator can lead to anti-flow. 
Secondly, NCQ scaling of $v_2$ is broken completely for particles ($\pi^+$, $K^+$, $K_S^0$, p, and $\Lambda$) and anti-particles ($\pi^-$, $K^-$) at $\sqrt{s_{NN}}$ = 3.2 GeV,
implying that the hadronic interactions are dominant at $\sqrt{s_{NN}}$ = 3.2 GeV and below. 
At last, collision energy dependence of $v_1$ slope ($dv_1/dy|_{y=0}$) and $p_T$-integrated $v_2$ at $\sqrt{s_{NN}}$ = 3.0 - 3.9 GeV are presented.
The $v_1$ slopes of all particles ($\pi^\pm$, $K_S^0$, and $\Lambda$) decrease in magnitude as collision energy increases. 
And the sign change in $v_2$ indicates that the change of out-of-plane to in-plane expansion happens between $\sqrt{s_{NN}}$ = 3.0 - 3.9 GeV.

\section*{Acknowledgement}
\label{acknow}
This work was supported in part by the National Key Research and Development Program of China (Nos. 2022YFA1604900 and 2020YFE0202002); the National Natural Science Foundation of China (Nos. 12175084); and Fundamental Research Funds for Central Universities (No. CCNU220N003).

\bibliography{ref}

\end{document}